\documentclass[aps,prl,showpacs,preprintnumbers,twocolumn,groupedaddress]{revtex4}
\usepackage{graphicx}
\usepackage{amsmath}
\usepackage{color}

\newcommand{\tr}{\mathrm{tr}}
\newcommand{\Retr}{\mathrm{Re \, tr}}

\begin{document}

\preprint{RIKEN-QHP-80}
\title{Lattice QCD in rotating frames}

\author{Arata~Yamamoto$^1$ and Yuji~Hirono$^{1,2,3}$}
\affiliation{
$^1$Theoretical Research Division, Nishina Center, RIKEN, Saitama 351-0198, Japan\\
$^2$Department of Physics, The University of Tokyo, Tokyo 113-0033, Japan\\
$^3$Department of Physics, Sophia University, Tokyo 102-8554, Japan}

\date{\today}

\begin{abstract}
We formulate lattice QCD in rotating frames to study the physics of QCD matter under rotation.
We construct the lattice QCD action with the rotational metric and apply it to the Monte Carlo simulation.
As the first application, we calculate the angular momenta of gluons and quarks in the rotating QCD vacuum.
This new framework is useful to analyze various rotation-related phenomena in QCD.
\end{abstract}


\maketitle

\emph{Introduction.}---Quantum Chromodynamics (QCD),  the fundamental theory of the strong
interaction, has been resulted in incredibly diverse phenomena.
Among them, QCD matter under rotation is of particular interest. 
For example, the cores of rapidly rotating compact stars are
expected to be in the superfluid or superconducting phase and are
threaded with topological vortices \cite{Forbes:2001gj}, which may play
an important role in pulsar glitches \cite{Anderson:1975}. 
In ultrarelativistic heavy-ion collisions performed at Relativistic Heavy Ion Collider (RHIC) or Large Hadron Collider (LHC), the
created quark-gluon plasma should have a finite angular momentum,
especially in off-central collisions. This would result in interesting
phenomena such as the chiral vortical effect \cite{Son:2009tf} or the
Kelvin-Helmholtz instability \cite{Csernai:2011qq}.
Also in low-energy nuclear physics, rotation generates
characteristic states of nucleus, e.g., rotational modes and high-spin
states \cite{deVoigt:1983zz}.
Thus, the analysis of the QCD matter under rotating environment is
an important phenomenological problem.

However, first principle calculations for the rotating QCD matter have
been elusive.
In numerical simulations under equilibrium conditions, such as the
lattice QCD simulation, it is not straightforward to generate a
rotating state.
The rotating state cannot be realized by adding external fields, since
the circulating velocity field, which characterizes the rotation, is
the matter field itself. 
One possible way is to rotate the reference frame.
A rest state in a rotating frame is equivalent to a rotating state in a
rest frame.
By performing the simulation in the rotating frame, we can study
the properties of the rotating matter.
In condensed matter physics, the simulation in the
rotating frame is successful in the analysis of rotating Bose-Einstein
condensates, in which interesting phenomena such as
vortex nucleation and the formation of a vortex lattice are simulated \cite{Fetter:2009}.

In this Letter, we formulate lattice QCD in rotating frames.
From phenomenological viewpoint, this framework opens a new possibility to the theoretical studies of QCD.
As mentioned above, there are many important rotation-related phenomena in hadron and nuclear physics.
Once this framework is formulated, we can tackle them by the first-principle calculation of QCD.
From theoretical viewpoint, this is one practical application of lattice QCD in a curved space-time, i.e., in general relativity.
Lattice QCD in moving frames has been discussed only in the case of the Lorentz boost, i.e., in special relativity \cite{Rummukainen:1995vs}.
Lattice gauge theory in a curved space-time has been discussed in the context of lattice quantum gravity \cite{Hamber:2009mt}.

\emph{In the continuum.}---In the standard lattice QCD simulation, we numerically perform the path integral with the Euclidean metric $g_{ij}={\rm diag}(1,1,1,1)$.
To formulate rotation in the Euclidean space-time, we need two operations: the Wick rotation $\tau = -it$ and the spatial rotation $\theta = \theta_{\rm rest}-\Omega t$.
There are two possibilities for the order of these two operations: the ``Minkowskian'' rotation $\Omega = -\partial \theta / \partial t$, which is defined as the spatial rotation before the Wick rotation, and the ``Euclidean'' rotation $\Omega = -\partial \theta / \partial \tau$, which is defined as the spatial rotation after the Wick rotation.
As is shown below, there is the sign problem in the Minkowskian rotation and no sign problem in the Euclidean rotation.
(This is similar to the case of external electric fields. There is the sign problem in the Minkowskian electric field $E_j = \partial A_j / \partial t - \partial A_0 / \partial x^j$ and no sign problem in the Euclidean electric field $E_j = \partial A_j / \partial \tau - \partial A_4 / \partial x^j$ \cite{Yamamoto:2012bd}.)
As long as the analytic continuation to the original Minkowski space-time is validated, these two rotations produce the same end result.
In this Letter, we apply the Euclidean rotation to the lattice QCD simulation.
In the following equations, the angular velocity of the Euclidean rotation is denoted by $\Omega$.
The corresponding equations with the Minkowskian rotation are obtained by replacing $\Omega \to i\Omega$.

We choose the Cartesian coordinate $x^{\mu} = (x,y,z,\tau)$
and perform the coordinate transformation to a rotating frame in the
Euclidean space-time.
The spatial coordinates in the rest and rotating frames are connected by
the relation 
$d x_{\rm rest}^i = dx^i - \epsilon^{ijk}\Omega_j x_k d\tau$, where $\Omega_j$ is the angular velocity vector \cite{Landau:1958}. 
By plugging this into the line element in the rest frame, 
$ds^2 = d\tau^2 + d \vec x_{\rm rest}^2$, we can read off the metric in the
rotating frame as 
\begin{eqnarray}
g_{\mu\nu} &=&
\begin{pmatrix}
1 & 0 & 0 & y\Omega \\
0 & 1 & 0 & -x\Omega \\
0 & 0 & 1 & 0 \\
y\Omega & -x\Omega & 0 & 1+r^2\Omega^2
\end{pmatrix}
,
\label{eqg}
\end{eqnarray}
where we chose $z$-axis as the rotation axis,  $\Omega_j =\delta_{jz} \Omega$ and $r \equiv \sqrt{x^2+y^2}$ is the distance from the rotation axis.
In the rotating frame, an observer feels the local Lorentz boost with the Lorentz factor $1/\sqrt{1+r^2\Omega^2}$.
In the Minkowskian rotation, for a well-defined coordinate patch, we must impose the condition $r\Omega <1=c$ (the light velocity), which means that the local velocity must be smaller than the light velocity.
In the Euclidean rotation, there is no such coordinate singularity but the local velocity should be small for the analytic continuation.

We start with the gluon and quark actions in a general curved space-time,
\begin{eqnarray}
S_G &=& \int d^4x \sqrt{\det g_{\alpha\beta}} \ \frac{1}{2g_{\rm YM}^2}
 g^{\mu\nu} g^{\rho\sigma} \tr F_{\mu\rho} F_{\nu\sigma} , \\
S_F &=& \int d^4x \sqrt{\det g_{\alpha\beta}} \ \bar{\psi}[\gamma^\mu (D_\mu-\Gamma_\mu) +m] \psi .
\end{eqnarray}
The covariant Dirac operator is constructed from the SU($N_c$) covariant derivative $D_\mu=\partial_\mu-iA_\mu$ and the spinor affine connection $\Gamma_\mu$. 
The connection is defined as
\begin{eqnarray}
\Gamma_\mu &=& -\frac{i}{4}\sigma^{ij} \omega_{\mu ij},\\
\sigma^{ij} &=& \frac{i}{2}(\gamma^i\gamma^j - \gamma^j\gamma^i),\\
\omega_{\mu ij} &=& g_{\alpha\beta} e^\alpha_i ( \partial_\mu e^\beta_j + \Gamma^\beta_{\nu \mu} e^\nu_j),
\end{eqnarray}
where $e_i^\mu$ is the vierbein and $\Gamma^\mu_{\alpha \beta}$ is the Christoffel symbol.
The Greek and Latin indices refer to the coordinate and tangent spaces, respectively.

By substituting the rotational metric (\ref{eqg}), the gluon action is
\begin{equation}
\begin{split}
S_G =& \int d^4x \ \frac{1}{g_{\rm YM}^2} \tr [ (1+r^2\Omega^2) F_{xy}F_{xy} \\
& + (1+y^2\Omega^2) F_{xz}F_{xz} + (1+x^2\Omega^2) F_{yz}F_{yz} \\
& + F_{x\tau}F_{x\tau} + F_{y\tau}F_{y\tau} + F_{z\tau}F_{z\tau} \\
& + 2y\Omega F_{xy}F_{y\tau} - 2x\Omega F_{yx}F_{x\tau} \\
& + 2y\Omega F_{xz}F_{z\tau} - 2x\Omega F_{yz}F_{z\tau} + 2xy \Omega^2 F_{xz}F_{zy} ].
\label{eqSG2}
\end{split}
\end{equation}
As an effect of rotation, the gluon action includes the $O(\Omega)$ terms which break parity and time-reversal symmetry and the $O(\Omega^2)$ terms which do not break them.
The covariant Dirac operator depends on the choice of the vierbein.
We choose the vierbein
\begin{eqnarray}
&& e^x_1 = e^y_2 = e^z_3 = e^\tau_4 = 1,\\
&& e^x_4 = -y\Omega, \quad e^y_4 = x\Omega,
\end{eqnarray}
and $e^\mu_i = 0$ for other components.
In this choice, the quark action is
\begin{equation}
\begin{split}
S_F =& \int d^4x \ \bar{\psi} \bigg[ \gamma^x D_x + \gamma^y D_y + \gamma^z D_z \\
& + \gamma^\tau \left( D_\tau +i \Omega \frac{\sigma^{12}}{2} \right) +m \bigg] \psi.
\label{eqSF2}
\end{split}
\end{equation}
The gamma matrices in the rotating frame are given as $\gamma^\mu =\gamma^i e_i^\mu $, i.e.,
\begin{eqnarray}
\gamma^x &=& \gamma^1 - y\Omega \gamma^4,\\
\gamma^y &=& \gamma^2 + x\Omega \gamma^4,\\
\gamma^z &=& \gamma^3,\\
\gamma^\tau &=& \gamma^4.
\end{eqnarray}
As a result of rotation, the Dirac operator includes the orbit-rotation coupling term $\gamma^\tau \Omega (xD_y-yD_x)$ and the spin-rotation coupling term $i \gamma^\tau \Omega \sigma^{12}/2$.

\begin{figure}[h]
\begin{center}
\includegraphics[scale=0.3]{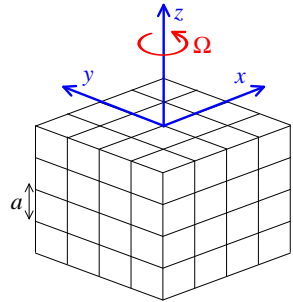}
\caption{\label{fig1}
Rotating lattice.
}
\end{center}
\end{figure}

\emph{On the lattice.}---We discretize the continuum actions (\ref{eqSG2}) and (\ref{eqSF2}) on the hypercubic lattice.
The schematic figure is shown in Fig.~\ref{fig1}.
The lattice spacing is $a$ and the total number of the lattice sites is $N_x \times N_y \times N_z \times N_\tau$.
The lattice is rotated around the $z$ axis.
In the $x$ and $y$ directions, we take the Dirichlet boundary conditions.
In the $z$ and $\tau$ directions, we take boundary conditions in the same manner as the usual lattice simulation.

On the lattice, the gluon field strength is constructed from the gauge invariant loops of the link variables $U_\mu(x)$.
The squared terms, e.g., $F_{xy}F_{xy}$, are constructed from the`` plaquette'' and the non-squared terms, e.g., $F_{xy}F_{y\tau}$, are constructed from the ``chair-type'' loop \cite{Iwasaki:1983ck}.
The lattice gluon action is
\begin{equation}
\begin{split}
S_G =& \sum_{x} \beta \bigg[ (1+r^2\Omega^2) \left(1-\frac{1}{N_c}\Retr \bar{U}_{xy} \right)  \\
& + (1+y^2\Omega^2) \left(1-\frac{1}{N_c}\Retr \bar{U}_{xz} \right) \\
& + (1+x^2\Omega^2) \left(1-\frac{1}{N_c}\Retr \bar{U}_{yz} \right) \\
& + 3-\frac{1}{N_c}\Retr \left( \bar{U}_{x\tau} + \bar{U}_{y\tau} + \bar{U}_{z\tau} \right) \\
& -\frac{1}{N_c}\Retr \big( y\Omega \bar{V}_{xy\tau} - x\Omega \bar{V}_{yx\tau} \\
& + y\Omega \bar{V}_{xz\tau} - x\Omega \bar{V}_{yz\tau} + xy \Omega^2 \bar{V}_{xzy} \big) \bigg].
\label{eqSG3}
\end{split}
\end{equation}
The bare lattice coupling is $\beta = 2N_c/g_{\rm YM}^2$.
For a local definition of the lattice field strength, we take the clover-type average of four plaquettes as
\begin{equation}
\bar{U}_{\mu\nu} = \frac{1}{4} \left( \parbox[c]{65pt}{\includegraphics[scale=0.35]{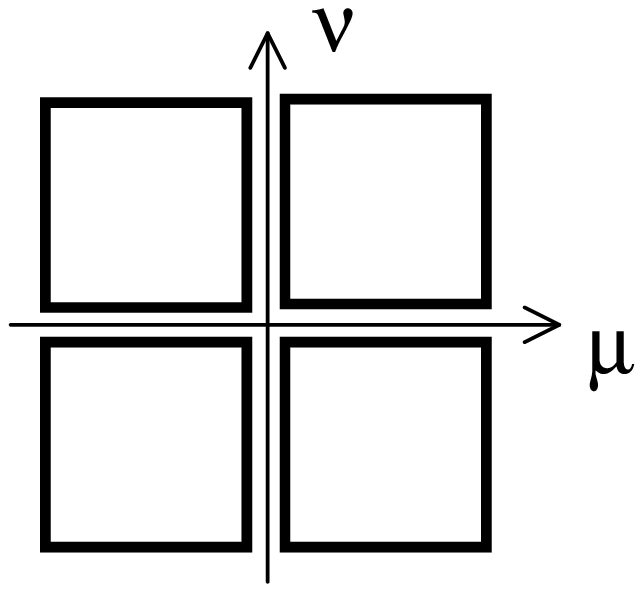}} \right).
\end{equation}
In a flat space-time, this average is redundant because it can be absorbed into the summation $\sum_x$ in the action.
However, in a curved space-time, this average is important because the coefficients depend on the space-time coordinate.
Similarly, we take the (anti-)symmetric average of eight chair-type loops as
\begin{equation}
\bar{V}_{\mu\nu\rho} = \frac{1}{8} \left( \parbox[c]{70pt}{\includegraphics[scale=0.35]{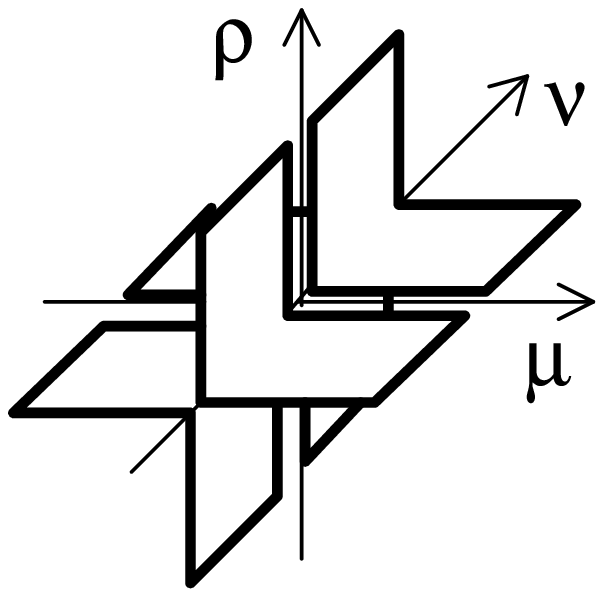}} - \parbox[c]{70pt}{\includegraphics[scale=0.35]{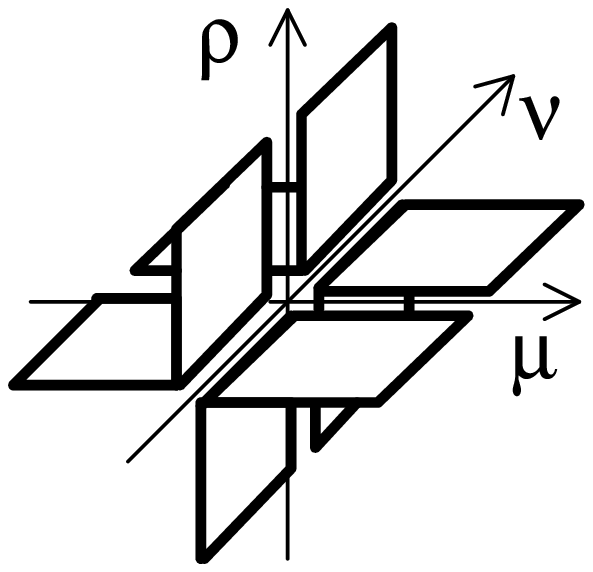}} \right).
\end{equation}
For the lattice quark action, we adopt the Wilson fermion with the gamma matrices in the rotating frame.
The spin-rotation coupling term is exponentiated as a chemical potential \cite{Hasenfratz:1983ba}.
The lattice quark action is
\begin{equation}
\begin{split}
S_F =& \sum_{x_1,x_2} \bar{\psi}(x_1) \bigg[ \delta_{x_1,x_2} \\
& - \kappa \bigg\{(1-\gamma^x) T_{x+} + (1+\gamma^x) T_{x-} \\
& + (1-\gamma^y) T_{y+} + (1+\gamma^y) T_{y-} \\
& + (1-\gamma^z) T_{z+} + (1+\gamma^z) T_{z -} \\
& + (1-\gamma^\tau) \exp \left(i a\Omega \frac{\sigma^{12}}{2} \right) T_{\tau +} \\
& + (1+\gamma^\tau) \exp \left(-i a\Omega \frac{\sigma^{12}}{2} \right) T_{\tau -} \bigg\} \bigg] \psi(x_2)
\end{split}
\label{eqSF3}
\end{equation}
with $T_{\mu +} \equiv U_{\mu}(x_1)\delta_{x_1+\hat{\mu},x_2}$ and $T_{\mu -} \equiv U^\dagger_{\mu}(x_2)\delta_{x_1-\hat{\mu},x_2}$.
The bare hopping parameter is $\kappa = 1/(2am+8)$.
In the continuum limit $a\to 0$, the lattice actions (\ref{eqSG3}) and (\ref{eqSF3}) correspond to the continuum actions (\ref{eqSG2}) and (\ref{eqSF2}), respectively.

In the Minkowskian rotation, the angular velocity is replaced as $\Omega \to i\Omega$.
In the gluon action (\ref{eqSG3}), the $O(\Omega)$ terms become pure imaginary numbers.
In the quark action (\ref{eqSF3}), the orbit-rotation coupling term becomes like an imaginary hopping term and the spin-rotation coupling term becomes like a chemical potential.
Since both of the gluon and quark actions
become complex, the Monte Carlo simulation severely suffers from the sign problem.
On the other hand, in the Euclidean rotation, the gluon and quark actions are real, and thus there is no sign problem.

We have formulated the hypercubic lattice, which is commonly used in most lattice simulations.
It is possible to formulate the cylindrical lattice in the cylindrical coordinate $x^{\mu} = (r,\theta,z,\tau)$.
An advantage of the cylindrical lattice is better rotational symmetry around the rotation axis.
However, since the action includes a singular metric factor $1/r$, the
region around the rotation axis must be removed to avoid this apparent singularity.

The angular velocity $\Omega$ might affect the renormalization, e.g., the physical scale.
This correction cannot be neglected when the angular velocity is large.
However, it is not trivial how to determine the physical scale in rotating frames.
Moreover, since the Lorentz symmetry and translational invariance are broken, the isotropy and the coordinate independence are no longer assured at the full quantum level.
(This is similar to the anisotropic lattice \cite{Hasenfratz:1981tw} and the coordinate-dependent lattice coupling \cite{Huang:1990jf}.)
In the following numerical simulation, we restrict the angular velocity only to small values, and do not discuss the problem of the renormalization.

\emph{Simulation.}---We performed the quenched SU(3) Monte Carlo simulation.
The lattice size is $N_x \times N_y \times N_z \times N_\tau=13 \times 13 \times 12 \times 12$.
The range of the $x$-$y$ plane is $x=[-6a,6a]$ and $y=[-6a,6a]$, and the position of the rotation axis is $(x,y)=(0,0)$.
We set the bare lattice coupling $\beta = 5.9$ and the bare hopping parameter is $\kappa = 0.1583$, where the lattice spacing is $a\simeq 0.10$ fm and the meson mass ratio is $m_\pi/m_\rho\simeq 0.59$ \cite{Aoki:2002fd}.

We analyze the angular momentum of the rotating QCD vacuum.
Rotation induces a finite vacuum expectation value of the angular momentum operator.
To understand the reason, let us recall a rotating classical particle.
The classical Lagrangian is $\mathcal{L} = mr^2\dot{\theta}^2_{\rm rest}/2 = mr^2(\dot{\theta}+\Omega)^2/2$, and it has a minimum at $\dot{\theta}=-\Omega$.
The classical solution has a finite angular momentum $J=mr^2\dot{\theta}=-mr^2\Omega$ in the rotating frame.
The negative sign means that the rest particle seems oppositely rotating from the rotating frame.
Similarly, in QCD, we can observe the rotating state with a finite angular momentum by generating the vacuum in a rotating frame.

We take the expectation value in the rotating vacuum,
\begin{equation}
J=\langle \hat{J} \rangle_{\Omega \ne 0},
\end{equation}
of the angular momentum density operator
\begin{equation}
\hat{J} \equiv \left. \frac{\partial \mathcal{L}}{\partial \Omega} \right|_{\Omega=0} ,
\end{equation}
where $\mathcal{L}$ is the Lagrangian density.
This angular momentum density operator coincides with the conserved Noether current in the flat space-time.
The gluon angular momentum density is
\begin{equation}
\begin{split}
J_G =& \bigg\langle \frac{1}{g_{\rm YM}^2} \tr [ 2y F_{xy}F_{y\tau} - 2x F_{yx}F_{x\tau} \\
& + 2y F_{xz}F_{z\tau} - 2x F_{yz}F_{z\tau}] \bigg\rangle.
\end{split}
\end{equation}
The fermion angular momentum density is decomposed into the orbital and spin angular momentum densities,
\begin{eqnarray}
J_F &=& J_{FL} + J_{FS}, \\
J_{FL} &=& \left\langle \bar{\psi} \gamma^\tau (xD_y-yD_x) \psi \right\rangle, \\
J_{FS} &=& \left\langle i \bar{\psi} \gamma^\tau \frac{\sigma^{12}}{2} \psi \right\rangle.
\end{eqnarray}
We discretize these operators in the same way as the lattice actions (\ref{eqSG3}) and (\ref{eqSF3}).

\begin{figure}[h]
\begin{center}
\includegraphics[scale=0.99]{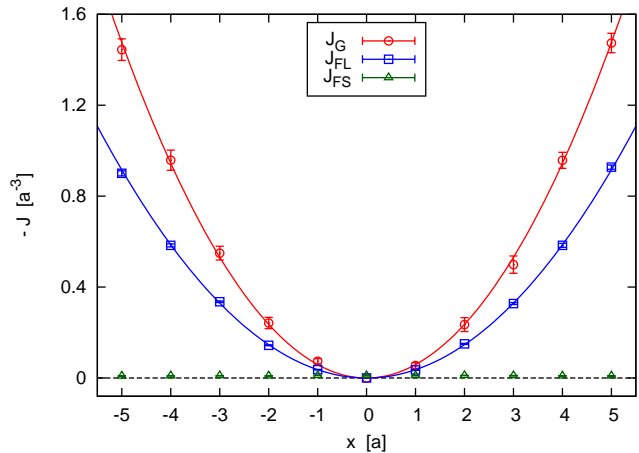}
\caption{\label{fig2}
Angular momentum density $J$ along the $x$ axis with the angular velocity $a\Omega = 0.06$.
The solid curves are quadratic fitting functions.
}
\end{center}
\end{figure}

\begin{figure}[h]
\begin{center}
\includegraphics[scale=0.99]{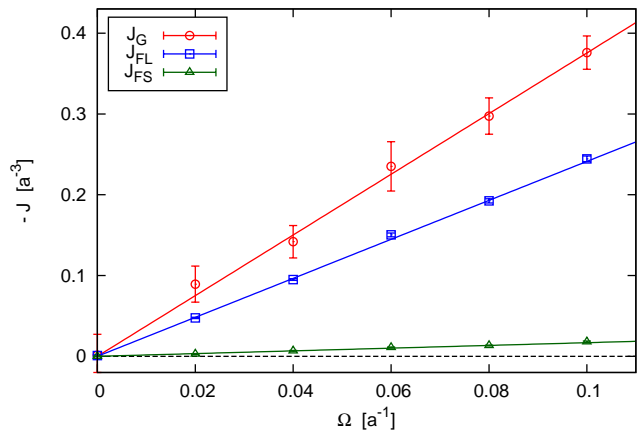}
\caption{\label{fig3}
Angular momentum density $J$ at $(x,y)=(2a,0)$ as a function of the angular velocity $\Omega$.
The solid curves are linear fitting functions.
}
\end{center}
\end{figure}

In Fig.~\ref{fig2}, we show the angular momentum density along the $x$ axis ($y=0$).
The angular velocity is fixed at a nonzero value $a\Omega = 0.06$.
As for $J_G$ and $J_{FL}$, the angular momentum density is a quadratic function of the distance from the rotation axis.
The spin angular momentum density $J_{FS}$ is small but nonzero, and it is independent of the distance.
In Fig.~\ref{fig3}, we show the angular momentum density measured at a certain point, $(x,y)=(2a,0)$, as a function of the angular velocity $\Omega$.
The angular momentum density increases linearly.
From fitting the data,
\begin{eqnarray}
J_{G}  &=& - (0.94 \pm 0.01) a^{-4} \times r^2\Omega, \\
J_{FL} &=& - (0.60 \pm 0.01) a^{-4} \times r^2\Omega, \\
J_{FS} &=& - (0.17 \pm 0.01) a^{-2} \times \Omega.
\end{eqnarray}
The coefficient in front of $\Omega$ is interpreted as the moment of inertia of the constituent in the QCD vacuum.
The functional form of $J_G$ and $J_{FL}$ can be intuitively understood
from the angular momentum of a classical particle, $J = -mr^2\Omega$. 
The numerical coefficients of $J_G$ and $J_{FL}$ are interpreted as the inertial mass densities of glueballs and quark-antiquark pairs, respectively.
The $r$-independence of $J_{FS}$ is a plausible result since the spin is an intrinsic angular momentum.
Note that these coefficients are unrenormalized and they depend on the renormalization scale and the quark mass.

\emph{Summary.}---We have formulated lattice QCD in rotating frames.
We have carried out its first Monte Carlo simulation to analyze the
angular momentum of the rotating QCD vacuum.
At least in the case of the Euclidean rotation, we can implement this
framework without technical difficulty.
By using this framework, we can study the rotating matter from
first principles.
There are many possible applications for QCD phenomenology, e.g.,
rotating hadrons, heavy-ion collisions, and rapidly rotating compact
stars. Moreover, this kind of simulation will be possible not only in QCD but
also in other field theories.

A.~Y.~is supported by the Special Postdoctoral Research Program of RIKEN.
Y.~H.~is supported by the Japan Society for the Promotion of Science for Young Scientists and by JSPS Strategic Young Researcher Overseas Visits Program for Accelerating Brain Circulation.
The numerical simulations were performed by using the RIKEN Integrated Cluster of Clusters (RICC) facility.

\end{document}